\documentclass[epsf,12pt]{article}
\usepackage{epsf}
\usepackage{amsmath}
\usepackage{amssymb}
\usepackage{graphicx}

\title{
Supersymmetry and Lie groups}
\author{Maciej Trzetrzelewski
\footnote{trzetrzelewski@th.if.uj.edu.pl }   \\
 \emph{M. Smoluchowski Institute of Physics, Jagiellonian
University} \\
\emph{Reymonta 4, 30-059 Krak\'ow, Poland}
}

\begin{document}
\date{}
\maketitle
\abstract{ We construct all vacuum states of $\mathcal{N}=2$ supersymmetric Yang-Mills
quantum mechanics ( for $SU(N)$ group ) and discuss their origin from the $SU(N)$ real cohomology.
\section{Introduction}

Supersymmetric quantum mechanics (SQM) provides an elegant and deep
connection between geometry and physics \cite{Witten}. It was
observed in early eighties by Witten that the number of the vacuum
states of $\mathcal{N}=2$ nonlinear sigma model on manifold $M$
coincides with the Betti numbers of the manifold. In this paper we
find an analogous correspondence in seemingly different models
namely $\mathcal{N}=2$ supersymmetric Yang-Mills quantum
mechanics (SYMQM) with $SU(N)$ gauge group. It turns out that the
number of vacuum states in sector with $n_F$ fermions coincides with
the $n_F$'th Betti number of $SU(N)$ manifold.  Such relations to
the topology, while well understood in the nonlinear sigma model,
were not, to our knowledge fully exhibited  for this system.

In  the following section, after determining all vacuum states of
$\mathcal{N}=2$ SYMQM, we identify them with nontrivial cocycles on $SU(N)$.
This serves as a mathematical proof that the number of vacuum
states coincides with the $SU(N)$ Betti numbers. We then conclude
that one can perform an identification, \`a la Witten, between the
left invariant forms on a Lie group and the fermion matrices
$\bar{\psi}=T_a\bar{\psi}_a$ where $T_a$ are matrix group
generators and $\bar{\psi}_a$ are fermion fields.

\section{A setup}

The $\mathcal{N}=2$ SYMQM involves real scalar
$\phi_a$, real gauge potential $A_a$ and the complex fermion field
$\psi_a$, $\bar{\psi}_a$ all in the adjoint representation of
$SU(N)$. The system can be obtained by the dimensional reduction
from  $D=1+1$, $\mathcal{N}=1$ gauge theory by rewriting the potential $A_{a
\ \mu}$, $\mu=0,1$ as $A_{a \ \mu}=(A_{a \ 0},A_{a \
1})\equiv(A_a, \phi_a)$.  The resulting lagrangian is \cite{Claudson}
\[
L=\frac{1}{2}(D_t \phi)_a(D_t \phi)_a+i\bar{\psi}_a(D_t
\psi)_a-igf_{abc}\bar{\psi}_a\phi_b \psi_c,
\]
where covariant derivatives are
\[
(D_t \phi)_a=\partial_t \phi_a-gf_{abc}A_b\phi_c, \ \ \ \ (D_t
\psi)_a=\partial_t \psi_a-gf_{abc}A_b\psi_c.
\]
The quantization of the system gives the hamiltonian
\[
H=\frac{1}{2}p_ap_a+g\phi_a G_a, \ \ \ \ G_a=i f_{abc}(\phi_b
p_c+\psi_b\bar{\psi}_c),  \ \ \ \ [G_a,G_b]=if_{abc}G_c,
\]
where $\phi_a$, $p_b$ and $\psi_a$,  $\bar{\psi}_b$ are conjugate
variables $[\phi_a,p_b]=i\delta_{ab}$,
$\{\psi_a,\bar{\psi}_b\}=i\delta_{ab}$. The Gauss law after dimensional
reduction becomes the singlet constraint on physical states $\mid s
\rangle$,  $G_a\mid s \rangle=0$. It follows now that in the
subspace of $SU(N)$ singlets the hamiltonian is very simple
\[
H=\frac{1}{2}p_ap_a=\frac{1}{2}\{ Q,\bar{Q} \},\ \ \ \
Q^2=\bar{Q}^2=0,   \ \ \ \ Q=\psi_a p_a, \ \ \ \
\bar{Q}=\bar{\psi}_a p_a,
\]
with the supercharges $Q, \bar{Q}$. Therefore the model is completely described
by the following equations
\[
\frac{1}{2}p_ap_a \mid s \rangle =E\mid s \rangle, \ \ \ \ G_a\mid s
\rangle=0.
\]

To make the system well defined we still have to take care of the
normalization of the vacuum $\mid v \rangle$: $H\mid v \rangle=0$ in
the purely bosonic sector $\psi \mid v \rangle=0$ ( it is worth
emphasizing that the vacuum state of this model differs from the
Fock vacuum $\mid 0 \rangle: \psi_a\mid 0 \rangle= (\phi_a+ip_a)\mid 0 \rangle=0$ ). Unfortunately if $\phi_a$'s are noncompact then this
requirement is not satisfied because the wave function of the vacuum $\mid v \rangle$
in coordinate representation is simply the constant function,
$\langle \phi \mid v  \rangle=1$. In fact the situation is even
worse since one can prove  \cite{33largeN} that there exist an
infinite number of polynomials
$P_k(Tr(\phi^2), \ldots, Tr(\phi^N))$ such that
\[
[p_a p_a,P_k]=2ik\partial_a P_k p_a \ \ \ \ \partial_a=\frac{\partial}{\partial \phi_a}.
\]
where we used the notation
\[
Tr(AB \ldots)=A_aB_b
\ldots Tr(T_a T_b \ldots),
\]
where $T_a$'s are $SU(N)$ generators in the
fundamental representation satisfying
\[
T_aT_b=\frac{2}{N}\delta_{ab}+(d_{abc}+if_{abc})T_c=\frac{2}{N}\delta_{ab}+\frac{1}{2}Tr(T_aT_bT_c)T_c.
\]
where $d_{abc}$, $f_{abc}$ are $SU(N)$ structure tensors.

Therefore, if $\mid v \rangle$ is the vacuum state then so is  $P_k\mid v \rangle$ hence
there are infinitely many vacua in purely bosonic sector.
Moreover, since the vacuum state is not normalizable the basic theorem in supersymmetry namely
\[
Q\mid v \rangle =0 ,\bar{Q}\mid v \rangle =0 \Longleftrightarrow H\mid v \rangle=0,
\]
does not have to hold anymore and it doesn't. To see this explicitly
we take the polynomial $Tr(\phi^3)$. We have
\[
p_ap_a Tr(\phi^3)\mid v \rangle=0,
\]
but
\[
\bar{Q} Tr(\phi^3)\mid v \rangle=-3iTr(\phi^2\bar{\psi})\mid v \rangle \ne 0.
\]

To remedy this ill situation we compactify  the coordinates $\phi_a
\in [0,1]$ and impose the periodicity condition on the wave function
$\Psi(\phi_a)=\Psi(\phi_a+1)$ in all fermion sectors. The lagrangian
is not invariant under the shift $\delta \phi_a =1$ but we also have
$\delta L = -ig\phi_aG_a$ therefore in the space of physical states
the compactification is properly imposed.
 The condition $\Psi(\phi_a)=\Psi(\phi_a+1)$  furnishes out
all additional solutions $P_k\mid v \rangle$ since they are not
periodic in $\phi_a$. Therefore there is only one vacuum $\mid v
\rangle$, in the sector with no fermions, which is now normalizable.

\section{The number of vacuum states}

Here we compute the number of vacua, of the model
described in previous section, in sectors with fermions. An arbitrary state with k fermions can be written as
\[
t_{i_1 \ldots i_k }(\phi)\bar{\psi}_{i_1} \ldots \bar{\psi}_{i_k} \mid v \rangle, \label{vac}
\]
where $t_{i_1 \ldots i_k }(\phi)$ are some functions depending on
$\phi_a$. One can proceed in two independent ways to count the vacua
a) by explicitly constructing the states and b) with use of the
representation theory.

We start with the first approach. The general form of the vacuum
state in the sector with $k$
fermions is
\begin{equation}
\mid v \rangle_k = t_{i_1 \ldots i_k }\bar{\psi}_{i_1} \ldots \bar{\psi}_{i_k} \mid v \rangle, \label{vac}
\end{equation}
where $t_{i_1 \ldots i_k}$ is $SU(N)$ invariant (due to the singlet
constraint)  tensor. There are no $\phi_a$s in (\ref{vac}) since any
appearance of them gives $\bar{Q}\mid v \rangle_k \ne 0 $. By the
same reason we act in (\ref{vac}) with fermion operators on the
bosonic vacuum $\mid v \rangle$ rather then on the Fock vacuum $\mid
0 \rangle$ since the wave function corresponding to $\mid 0 \rangle$
depends on $\phi_a$ i.e. $\langle \phi \mid 0 \rangle \propto
\exp(-Tr(\phi^2)/2)$.

Invariant tensors can be expressed as linear combination of products
of trace tensors $Tr(T_aT_b \ldots)$ therefore the following states
\[
Tr(\bar{\psi}^{2})^{i_2}\ldots Tr(\bar{\psi}^{N^2-1})^{i_{N^2-1}} \mid v \rangle,
\]
span the entire space of vacuum states in all fermion sectors.
Moreover, since fermions anticommute we have $Tr(\bar{\psi}^{2k})=0$
and $Tr(\bar{\psi}^{2k+1})^2=0$. Therefore we are left with the states
\[
Tr(\bar{\psi}^{3})^{i_3}Tr(\bar{\psi}^{5})^{i_5}\ldots
Tr(\bar{\psi}^{N^2-1})^{i_{N^2-1}} \mid v \rangle, \ \ \ \ i_k=0,1, \ \ \ \ N^2-1 \ \ \
\hbox{odd},
\]
\[
Tr(\bar{\psi}^{3})^{i_3}Tr(\bar{\psi}^{5})^{i_5}\ldots
Tr(\bar{\psi}^{N^2-2})^{i_{N^2-2}} \mid v \rangle, \ \ \ \ i_k=0,1, \ \ \ \ N^2-2 \ \ \
\hbox{odd}.
\]
They can be further reduced due to the following fact. The
multiplication law for $T_a$'s gives us
\[
\bar{\psi}\bar{\psi}=\frac{1}{2}Tr(\bar{\psi}\bar{\psi}T_a)T_a, \ \ \ \
\bar{\psi}=\bar{\psi}_aT_a,
\]
therefore
\[
Tr(\bar{\psi}^{2n+1})= \frac{1}{2^n}Tr(\bar{\psi}\bar{\psi}T_{a_1})\ldots
Tr(\bar{\psi}\bar{\psi}T_{a_n})Tr(T_{a_1}\ldots T_{a_n} \bar{\psi}).
\]
Since operators $Tr(\bar{\psi}\bar{\psi}T_{a_k})$ commute with each
other we may symmetrize over indices

\[
Tr(\bar{\psi}^{2n+1})=\frac{1}{2^nn!} Tr(\bar{\psi}\bar{\psi}T_{a_1})\ldots
Tr(\bar{\psi}\bar{\psi}T_{a_n})Tr(T_{(a_1}\ldots T_{a_n)} \bar{\psi}).
\]
Generators $T_a$ are $N \times N$ matrices therefore  according to
Cayley-Hamilton theorem if $n \ge N$ then the matrix $T_{(a_1}\ldots
T_{a_n)}$ can be expressed as a linear combination of products of
matrices $T_{(a_1}\ldots T_{a_k)}$, $k<n$.
This implies that operators $Tr(\bar{\psi}^{2n+1})$, $n \ge N$
 can be expressed as a linear combination of products of
operators $Tr(\bar{\psi}^{2n+1})$, $n<N$.

Therefore we are left with the following vacuum states
\[
\mid v \rangle_k =Tr(\bar{\psi}^{3})^{i_3}Tr(\bar{\psi}^{5})^{i_5}\ldots
Tr(\bar{\psi}^{2N-1})^{i_{2N-1}} \mid v \rangle, \ \ \ \ i_k=0,1.
\]
Let us denote the number of vacua in sector with $k$ fermions by
$b_k$. If follows that the generating polynomial for $b_k$'s is
\begin{equation}
P(t)=\sum_{i=0}^{N^2-1} b_i t^i=(1+t^3)(1+t^5)\ldots(1+t^{2N-1}), \label{1}
\end{equation}
which is exactly the Poincar\'e polynomial for the $SU(N)$ manifold  (
the collection of Poincar\'e polynomials for other compact
semisimple Lie groups can be found in, e.g. \cite{Boya} or \cite{Az}
). This result is somewhat puzzling since it is not entirely clear why the
$\mathcal{N}=2$, SYMQM should have any topological interpretation
analogous to nonlinear sigma models. Before we give the answer to
this puzzle we will present yet another derivation of the above
result with use of representation theory.

Let $F_a$ be the vector space spanned  by operators $\bar{\psi}_a$.
The fermions are in the adjoint representation of $SU(N)$ which we
denote by $R$. It follows that the state with $k$ fermions belongs
to the tensor product $V=Alt(\otimes_{a=1}^{k} {F_a})$ where $Alt$
means the antisymmetrization of the tensor product. The number of
independent vacua $b_k$ is simply the number of $SU(N)$ singlets in
$V$ therefore
\[
b_k= \int
d \mu_{\textbf{SU(N)}}  \chi_{Alt}^{[k]}(R),  \label{betti}
\]
where the $SU(N)$ invariant  measure   $d \mu_{\textbf{SU(N)}}$, the
antisymmetric power of $R$, $\chi_{Alt}^{[a_F]}(R)$ and the
characters $\chi(R)$  are listed in the Appendix where we also prove
that the generating function (\ref{1}) has the following integral
representation

\begin{equation}
P(t) =\frac{1}{N!}(1-t)^{N-1}
\int_{[0,2\pi]^N}\prod_i\frac{d\alpha_i}{2\pi}\delta(\alpha_N)
\prod_{i \ne j}(1-\frac{z_i}{z_j})(1-t\frac{z_i}{z_j}),
\end{equation}
where $z_j=e^{i \alpha_j}$ and $\alpha_j=[0,2\pi]$.
 For given $N$ the above integral can be
evaluated it in terms of residues and it reproduces (\ref{1}) as it
should.

The connection with the group theory becomes even more evident  if
we realize that the vacuum states  correspond to the non-trivial
cocycles for the $SU(N)$ real cohomology. To be more specific, ( see e.g.
\cite{Az} for more details ), consider a basis ${X_1\mid_{e},\ldots,
X_{N^2-1}\mid_e}$ of tangent space $T_e SU(N)$, where $e$ is the
identity element of $SU(N)$. One can define the basis of linearly
independent left-invariant vector fields ${X_1,\ldots, X_{N^2-1}}$
at each point $g \in SU(N)$ by $X_a\mid_g=L_{g^*}X_a\mid_e$,  where
$L_{g^*}$ is the $SU(N)$ automorphism induced by the left
translation $L_g$. Let $\theta^a$ be dual to $X_a|_g$ and let us
consider the Maurer-Cartan form $\theta(g)=\theta^a(g)X_{a}$. Due to
the Maurer-Cartan equation $d\theta= -\theta\wedge\theta$ the following form
\[
\Omega^n(g)=Tr(  \underbrace{ \theta
\wedge \ldots \wedge \theta}_{n} ), \ \ \ \ n \hbox{ - odd},
\]
is closed but not
exact, therefore it defines the well known Chevalley-Eilenberg
$n$-cocycle of $SU(N)$. We recognize that the cocycles
$\Omega^n(g)$ correspond to operators $Tr(\bar{\psi}^n)$ needed
to construct the vacuum states \footnote{ The
explicit construction of coordinates of $\Omega^n(g)$ and their
properties are discussed in \cite{Az1}. }. Therefore, the number of vacuum states in the
sector with $n_F$ fermions coincides with the number of independent,
closed, not exact, cocycles that one can build on $SU(N)$, i.e. the
$n_F$'th Betti number.

One can continue the analogy between   $Tr({\bar{\psi}}^n)$ and
$\Omega^n(g)$  by identifying $\bar{\psi}$ with the Maurer-Cartan
1-form $\theta$ and the multiplication of fermions
$\bar{\psi}\bar{\psi}$ with the wedge product $\theta \wedge
\theta$. The vacuum state in the bosonic sector is then a constant
0-form equal 1.

\section{Summary}

In this paper we investigated the vacuum structure
of $\mathcal{N}=2$, SYMQM. It turns out that the vacua reveal a topological information of the gauge group
considered. If we look at this system just regarding the hamiltonian
and the singlet constraint it is unclear why there should be any
such information. The necessity of the singlet constraint can be
seen from the following argument. One could consider the
hamiltonian $H=\frac{1}{2}p_ap_a$
 without the constraint and the system still remains
supersymmetric only this time the number of vacuum states does not
coincide with Betti numbers.\footnote{This time the number of vacuum
states and the corresponding generating polynomials are
\[
b_{n_F}= \binom{N^2-1}{n_{F}}, \ \ \ \  P(t)=(1+t)^{N^2-1}.
\]  }

It turns out that  the construction of
vacuum states coincides with the well known construction of
non-trivial cocycles on $SU(N)$. The coordinates of those cocycles
are group invariant tensors which is precisely the requirement
coming from the Gauss law.

We focused entirely on the $SU(N)$ group, however the  case of other
Lie groups is analogous although it is necessary for the Lie group
to be compact.

\section{Acknowledgments}
I thank R. Janik, G. Veneziano, P. Di Vechia and J. Wosiek for
discussions. I also thank the referee for useful comments.

This work was supported by the grant of Polish Ministry of
Science and Education no. P03B 024 27 ( 2004 - 2007 ) and N202 044
31/2444 ( 2006-2007 ) and the  Jagiellonian University Estreicher
foundation.

\section{Appendix A}
Here we give the conventions used in section 3 and prove Eqn. (\ref{1}).
The conventions we use can be found in \cite{Itz}. The
$SU(N)$ invariant, normalized,  measure is
\[
d\mu_{SU(N)} =
\frac{1}{N!}\prod_{i=1}^{N}\frac{d\alpha_i}{2\pi}\delta_P(\sum_{i=1}^{N}
\alpha_i) \mid M \mid^2, \ \ \ \ \alpha_i \in [0,2\pi],
\]
where $\delta_P$ is a periodic Dirac delta with period $2\pi$
\[
\delta_P(x)=\sum_k\delta(x-2\pi k),
\]
the measure factor $M$ is given by Vandermonde determinant
\[
M=Det(z_j^{(N-i)})=\prod_{i<j} (z_i-z_j), \ \ \ \ z_j=e^{i \alpha_j},
\]
and $\chi_{Alt}^{[a_F]}(R)$ is the antisymmetric power of $R$
given by Frobenius formula
\begin{equation}
\chi_{Alt}^{[a_F]}(R)=\sum_{\sum_k k i_k=n_F} (-1)^{\sum_k i_k}
\prod_{k=1}^{n_B}
\frac{1}{i_k!}\frac{\chi^{i_k}(R^k)}{k^{i_k}}, \label{fro}
\end{equation}
where the characters $\chi$ are given by
Weyl determinant formula
\[
\chi(R)\equiv\chi(\{\alpha_i\}_{i=1}^N)=
\frac{Det(z_j^{(N-i+l_i)})}{Det(z_j^{(N-i)})}, \ \ \ \ \chi(R^k)=
\chi(\{k\alpha_i\}_{i=1}^N).
\]
The numbers $l_i$ enumerate the representation in which the
character is computed. In our case it is the adjoint
representation of $SU(N)$ therefore
$(l_1,l_2,\ldots,l_N)=(2,1,\ldots,1,0)$ . In this representation
the characters simplify into
\begin{equation}
\chi_{SU(N)}(\{\alpha_i\})=\sum_{ i,j }\frac{z_i}{z_j} -1 \label{char}.
\end{equation}
Since for  $SU(N)$ we have $\chi_{Alt}(R^{k})=0$ when $k>N^2-1$
we can write the generating function (2) as an infinite sum
\begin{equation}
P(t)=\sum_{i=0}^{\infty} b_i t^i.  \label{poly}
\end{equation}
Substituting  (\ref{betti}), (\ref{fro}), (\ref{char})  to (\ref{poly}) we obtain (after some manipulations)
\[
P(t)=\int
d \mu_{\textbf{SU(N)}} \exp \left( \sum_{k=1}^{\infty}
\frac{t^k}{k}\chi(\{k\alpha_i\}_{i=1}^N)\right).
\]
Using the formula for characters and the measure we obtain
\[
P(t)=\frac{1}{N!}\int_{[0,2\pi]^{N-1}}
\delta(\alpha_N)\prod_i\frac{d\alpha_i}{2\pi}  \prod_{i \ne
j}(1-\frac{z_i}{z_j})\prod_{i,j}(1-t\frac{z_i}{z_j}),
\]
where we also changed variables  $z_i \rightarrow
z_i/\prod_{j=1}^N z_j$, $z_N\rightarrow \prod_{j=1}^N z_j $.

\end{document}